\input harvmac

\input amssym.def
\input amssym

\def\frac#1#2{{#1\over#2}}

\def\exp{{\rm exp}}

\def\s{\sigma}

\def\Tr{{\rm Tr}}
\writedefs

\lref\Kel{Keldysh, L.V.: Diagram technique for nonequilibrium 
processes. Sov. Phys. JETP {\bf 20}, 1018-1026 (1965)}

\lref\DDV{Destri., C. and de Vega, H.J.:
Unified approach to thermodynamic Bethe Ansatz 
and finite size corrections
for lattice models and field theories. 
Nucl. Phys. {\bf B438}, 413-454 (1995)}

\lref\Wei{Weiss, U.: Low-temperature conduction and DC current in a
quantum wire  with impurity. Sol. St. Comm. {\bf 100}, 281-285 (1996)}

\lref\KBP{Kl\"umper, A., Batchelor, M.T. and Pearce P.A.: Central
charges of the 6- and 19-vertex models with twisted boundary
conditions.
J. Phys A: Math. Gen. {\bf 24}, 3111-3133 (1991)}
 
\lref\Fen{Fendley, P., Ludwig, A.W.W.  and Saleur, H.:
Exact non-equilibrium transport through point contacts in
quantum wires and fractional quantum Hall devices.
Phys. Rev. {\bf B52},
8934-8950 (1995)}

\lref\Zams{ Zamolodchikov, Al.B.:
On the TBA equations for reflectionless ADE
scattering theories.
Phys. Lett. {\bf B253}, 391 (1991)}

\lref\Fey{Feynman R.P. and Vernon, F.L.: The theory of a
general quantum system interacting with
a linear dissipative system.
Ann. Phys. (NY) {\bf 24}, 118-173 (1963) }

\lref\BLZ{Bazhanov, V.V., Lukyanov, S.L. and Zamolodchikov, A.B.:
Integrable structure of conformal field theory,
quantum KdV theory and
thermodynamic Bethe Ansatz.
Commun. Math. Phys. {\bf 177}, 381-398 (1996)}

\lref\FSLSN{Fendley, P., Ludwig, A.W.W. and Saleur H.:
Exact conductance through point contacts in
the $\nu=1/3$ fractional quantum Hall effects.
Phys. Rev. Lett. {\bf 74}, 3005-3008 (1995) \semi
Fendley, P., Ludwig, A.W.W.  and Saleur, H.:
Exact non-equilibrium transport through point contacts in
quantum wires and fractional quantum Hall devices.
Phys. Rev. {\bf B52},
8934-8950 (1995)}

\lref\Baxn{Baxter, R.J.:
Eight-vertex model
in lattice statistics and
one-dimensional anisotropic
Heisenberg chain\semi
1. Some fundamental 
eigenvectors. Ann. Phys. (N.Y.) {\bf 76}, 1-24 (1973)\semi
2. Equivalence to
a generalized ice-type model. Ann. Phys. (N.Y.)
{\bf 76}, 25-47 (1973)\semi
3. Eigenvectors of the transfer matrix and Hamiltonian.
Ann. Phys. (N.Y.) {\bf 76}, 48-71 (1973)}

\lref\weiss{Weiss, U., Egger, R. and Sassetti, M.: Low-
temperature nonequilibrium transport in a Luttinger liquid.
Phys. Rev. {\bf B52}, 16707-16719 (1995)}

\lref\BLZZ{Bazhanov, V.V., Lukyanov, S.L. and Zamolodchikov, A.B.:
Integrable structure of conformal field theory.
2. Q-operator
and DDV equation. Commun. Math. Phys. {\bf190}, 247-278 (1997)}

\lref\BLZZZ{Bazhanov, V.V., Lukyanov, S.L. and Zamolodchikov, A.B.:
Integrable structure of conformal field theory. The Yang-Baxter
relation. Preprint RU-98-14 (1998) (\# hep-th 9805008)} 

\lref\BLZZZZ{Bazhanov, V.V., Lukyanov, S.L. and Zamolodchikov, A.B.:
Integrable quantum field theories in finite volume: Excited state 
energies. 
Nucl. Phys. {\bf B489}, 487-531 (1997)} 

\lref\Kane{Kane, C.L. and Fisher, M.P.A.:
Transmission through barriers and resonant tunneling in
an interacting one-dimensional electron gas.
Phys. Rev.  {\bf B46}, 15233-15262 (1992)} 

\lref\Moon{Moon, K., Yi, H., Kane, C.L., Girvin, S.M. and
Fisher, M.P.A.: Resonant tunneling between quantum Hall
edge states. Phys. Rev. Lett. {\bf 71}, 4381-4384 (1993) }

\lref\SalF{Fendley, P., Lesage, F. and Saleur H.:
A unified framework for the Kondo problem and
for an impurity in a Luttinger liquid.
J. Stat. Phys. {\bf 85}, 211-249 (1996)}

\lref\Schmid{Schmid, A.:
Diffusion and localization in a dissipative
quantum system.
Phys. Rev. Lett. {\bf 51}, 1506-1509 (1983)}

\lref\leggett{Caldeira, A.O.
and Leggett, A.J.: Influence
of dissipation on quantum tunneling in
macroscopic systems. Phys. Rev. Lett. 
{\bf 46}, 211-214 (1981)\semi
Caldeira, A.O.
and Leggett, A.J.:
Path integral approach to quantum
Brownian motion.
Physica  {\bf A121}, 587-616 (1983)}

\lref\Callan{Callan, C.G. and Thorlacius, L.:
Open string theory as dissipative quantum 
mechanics. Nucl. Phys. {\bf B329}, 117-138 (1990)}

\lref\Fisher{Fisher, M.P.A. and Zwerger, W.:
Quantum Brownian motion in a
periodic potential.  Phys. Rev.
{\bf B32}, 6190-6206 (1985)}

\Title{\vbox{\baselineskip12pt\hbox{RU-98-51}
\hbox{hep-th/9812091}}}
{\vbox{\centerline{On nonequilibrium states}
{\centerline{}} 
{\centerline{in  QFT model with boundary interaction}}
\vskip3pt\centerline{  }}}
\centerline{Vladimir V. Bazhanov$^1$,
Sergei L. Lukyanov$^{2}$}
\centerline{ 
and Alexander B. Zamolodchikov$^{2}$ }
\centerline{ }
\centerline{$^1$Department of 
Theoretical Physics and Center of Mathematics}
\centerline{and its Applications, IAS, Australian National University, }
\centerline{Canberra, ACT 0200, Australia}
\centerline{ }
\centerline{$^2$Department of Physics and Astronomy,}
\centerline{Rutgers University, Piscataway, NJ 08855-049, USA}
\centerline{and}
\centerline{L.D. Landau Institute for Theoretical Physics,}
\centerline{Chernogolovka, 142432, Russia}

\centerline
\centerline
\centerline{{\bf Abstract}}
\centerline{}

We prove that certain nonequilibrium expectation values in the boundary
sine-Gordon model coincide with associated equilibrium-state expectation
values in the systems which differ from the boundary sine-Gordon in that
certain extra boundary degrees of freedom ($q$-oscillators) are
added. Applications of this result to actual calculation of
nonequilibrium characteristics of the boundary sine-Gordon model are
also discussed. 

\Date{December, 98}
\vfil
\eject

\newsec{Introduction}

In this paper we study the 
so called boundary sine-Gordon model with zero bulk 
mass (referred below as BSG). Its action is 
\eqn\bsga{ 
{\cal A}_{BSG} = {1\over {4 \pi g}}\int_{-\infty}^{\infty}dt 
\int_{-\infty}^{0} dx\, \big(\, \Phi_{t}^2 - \Phi_{x}^2\, \big) + 
{\kappa\over g}\int_{-\infty}^{\infty}dt\, \cos\big(\Phi_{B}+Vt\big)\ .} 
Here $\Phi = \Phi(x,t)$ is a scalar field defined on a half line 
$-\infty < x \leq 0$, $\Phi_{B} \equiv \Phi(0,t)$ is its boundary
value, $\Phi_t=\partial_t \Phi$,\ $\Phi_x=\partial_x \Phi$\ 
and  $g$, $\kappa$ and $V$ are parameters. The way $g$ enters
\bsga\ allows one to interpret it as a quantum parameter, because it
always appears in the combination $g \hbar$; in what follows we
set $\hbar=1$. In the quantum theory the parameter $\kappa$ carries the
dimension \foot{We assume here that the
normalization of the boundary field $\cos(\Phi_B +Vt)$ is fixed by the
condition $\langle\,  \cos(\Phi_B (t) + Vt)\,\cos(\Phi_B (t') +
Vt')\, \rangle_{BSG} \to 2^{-1}\, \big(\, 
 i(t-t')+0\big)^{-2 g}$ as $0<t-t' \to 0$.} of $[mass]^{1-g}$. The
boundary interaction in \bsga\ contains explicit time dependence through
the term $Vt$. Although this time dependence can be eliminated by a simple
change of the field variables $\Phi \to \Phi - Vt$ (which instead brings
in the term linear in $\Phi_{t}$), the above form \bsga\ is more
convenient for our analysis. Below we almost always assume that $V> 0$,
more generally we will consider complex $V$ with $\Re e\, V > 0$. 
For future references let us write down the Hamiltonian corresponding 
to \bsga,
\eqn\bsgh{
{\bf H}_{BSG} = {1\over {4\pi g}}\int_{-\infty}^{0}dx\, \big(\, 
{\Pi}^2 + 
\Phi_{x}^2\,  \big)- {\kappa\over g}\cos\big(\Phi_{B} + Vt\big)\ .} 
Here $\Phi(x),\,  {\Pi}(x)$ are field operators obeying canonical 
commutation relations 
\eqn\ccr{
\big[\Pi(x)\, , \Phi(x')\big] = -2\pi i g\  \delta (x-x') } 
and again $\Phi_{B} \equiv \Phi(0)$. 

At a nonzero $V$ and a temperature $T$ the system \bsgh\ develops a
stationary nonequilibrium state 
which can be thought of as the result of an
infinite time evolution of the equilibrium state of the corresponding
``free'' system, with the interaction term
(the last term in \bsgh) adiabatically switched on. We will denote 
$\langle\,{\bf A}\, \rangle_{BSG}$  the
expectation value of an observable ${\bf A}$ over this 
nonequilibrium stationary state. 

Besides being an interesting model of Quantum Field Theory on its own, 
the theory finds important applications in other branches of
physics. As explained in\ \refs{\leggett, \Callan}, the model \bsga\
describes a quantum 
particle with the coordinate $X=\Phi_B+Vt$ and the potential energy 
$-2\pi\kappa\, \cos( X )$, interacting with dissipative environment,
the bulk part of the field $\Phi$ playing the role of the latter
(see also \refs{\Schmid,\Fisher}). In
this case $V$ is interpreted as an external driving force. The model
\bsga\ is also believed to describe an electric current through a point 
contact in the quantum Hall system\ \refs{\Kane}; 
in this case $V$ is proportional to the 
voltage drop across the contact. In all cases the quantities of interest
are the correlation functions of the boundary fields
\eqn\vpm{
{\bf V}_{+}(t) = e^{i\Phi_B (t)}\ e^{iVt}\, , 
\qquad {\bf V}_{-}(t) = e^{-i\Phi_B (t)}\ e^{-iVt}\ .}
Here $\Phi_B (t) ={\bf S}^{-1}(t)\,
\Phi_B\, {\bf S}(t)$, and ${\bf S}(t)$ is the time
evolution operator corresponding to the Hamiltonian\ \bsgh. 
We will be 
particularly interested in the expectation
values $\langle\,{\bf V}_{\pm}\, \rangle_{BSG}$.

The main goal of this paper is to show that the expectation values 
$\langle\,{\bf V}_{\pm}\, \rangle_{BSG}$ (and indeed some
more general correlation functions of
\bsga) coincide with {\it equilibrium} expectation values of
certain operators in 
a system which differs from \bsgh\ in that it involves certain additional 
boundary degree of freedom. Namely, let us define a Hamiltonian
\eqn\hplus{
{\bf H}_{+} = {1\over {4\pi g}}\int_{-\infty}^{0}dx\, 
\big(\, \Pi^2 + 
\Phi_{x}^2\,  \big) -V\,{\bf h} - 
{\kappa\over{2 g}}\ \big(\, 
{\bf a}_{-}\, e^{i\Phi_B}+{\bf a}_{+}\, e^{-i\Phi_{B}}\, \big)\ ,
}
where $\Phi(x),\,  \Pi(x)$ are again the Bose field operators
obeying the same commutation relations \ccr\ as in \bsgh, and the
operators ${\bf h},\, {\bf a}_{+},\, 
{\bf  a}_{-}$ commute with $\Phi(x),\, \Pi(x)$ 
and form among themselves the so called ``$q$-oscillator algebra'', i.e.
\eqn\qosc{
\big[{\bf h}, {\bf a}_{\pm}\big]=\pm  {\bf a}_{\pm}\, ; 
\quad q\,  {\bf a}_{+} {\bf a}_{-} - q^{-1}\, {\bf a}_{-} {\bf a}_{+} =
q - q^{-1} 
}
with
\eqn\qdef{
q=e^{i\pi g}\ .
}
Let $\rho_{+}$ be some representation of \qosc\ such that the spectrum 
of $\rho_{+} ({\bf h})$ is real and bounded from above. The 
Hamiltonian \hplus\ acts in the space
\eqn\spaceplus{
{\cal H}_{+}={\cal F}\otimes \rho_{+}\ ,
}
where ${\cal F}$ is the space of states of the Bose field representing
the commutation relations \ccr, and for $V>0$ this Hamiltonian is
bounded from below. Then, for $V>0$  the system \hplus\ has a thermal
equilibrium state described by the standard density matrix 
\eqn\hibbs{
{\bf P}_{+}=Z_{+}^{-1}(\kappa, V)\ e^{ -R {\bf H}_{+}}\ ,
}
where $R$ is  proportional to the inverse temperature,
$$R=g/T\ ,$$
and $Z_{+}(\kappa, V) =
{\rm Tr}_{{\cal H}_{+}}\big[\,e^{-R{\bf H}_{+}}\, \big]$ is the
corresponding  partition
function. Here and below we treat $g$ as a constant, and therefore
we do not include it in the list of arguments of $Z_{+}$.
Let us denote $\langle\,{\bf A}\, \rangle_{+}$ the expectation 
value of an observable ${\bf A}$ over 
this thermal equilibrium state. We will show that
\eqn\main{
\langle\, {\bf V_{+}}\, \rangle_{BSG} = 
\langle\, {\bf W}_{+}\, \rangle_{+} = 
\langle\, {\bf  W_{-}}\, \rangle_{+}\ ,
}
where
\eqn\wdef{
{\bf W}_{+}={\bf a}_{-}\, e^{i\Phi_B}\, ;
\quad{\bf W}_{-} = {\bf a}_{+}\,  e^{-i\Phi_{B}}\ .
}
While the second equality in \main\ is a simple property of the
equilibrium state \hibbs\ (see Sect.3), 
the relation between the nonequilibrium
and equilibrium expectation values in \main\ looks rather unusual and
suggestive. In fact, the Eq.\main\ is a particular case of a more general
relation. Namely, the correlation functions in \bsgh\ involving any 
number of the Heisenberg operators ${\bf V}_{+}(t)$ 
(but not ${\bf V}_{-}(t)$) coincide
with  the corresponding equilibrium-state correlation functions of the
operators ${\bf W}_{+}(t)$ in \hplus. 

Similarly, one can relate the correlation functions which involve
the operators ${\bf V}_{-}(t)$ (but not ${\bf V}_{+}(t)$) of \bsgh\ 
to certain
equilibrium-state correlation functions. Let $\rho_{-}$ be any 
representation of \qosc\ such that the spectrum of
$\rho_{-}({\bf h})$ is bounded from below. Consider the Hamiltonian 
\eqn\hminus{
{\bf H}_{-} ={1\over {4\pi g}}\int_{-\infty}^{0}dx\, 
\big(\, {\Pi}^2 +
\Phi_{x}^2\,  \big)  + V\,{\bf h} -
{\kappa\over{2 g}}\ \big(\, {\bf a}_{+}\, e^{i\Phi_B}
 + {\bf a}_{-}\, e^{-i\Phi_{B}}\, \big)
}
acting in ${\cal F}\otimes \rho_{-}$, and associated equilibrium state
density matrix
\eqn\hibbsminus{
{\bf P}_{-}=Z_{-}^{-1}(\kappa, V)\ e^{ -R {\bf H}_{-}}\ ,
}
which is well defined for $\Re e\, V>0$. Then
\eqn\mainminus{
\langle\,  {\bf V}_{-}\, \rangle_{BSG} = 
\langle\, {\bf {\tilde W}}_{-}\, \rangle_{-} =
\langle\, {\bf {\tilde W}}_{+}\, \rangle_{-}\ ,
}
where $\langle\,  {\bf A}\,  \rangle_{-}$ stands 
for the expectation value of ${\bf A}$ over the
equilibrium state \hibbsminus, and
\eqn\wwdef{
{\bf {\tilde W}}_{+}={\bf a}_{+}\, e^{i\Phi_B}
\, ; \quad {\bf {\tilde W}}_{-}={\bf a}_{-}\, e^{-i\Phi_B}\ .
}
Although the Hamiltonian \hminus\ is related to \hplus\ through a simple
change $\Phi \to -\Phi, \ V \to -V$, we treat it as a distinct one
because we always imply $\Re e\, V > 0$.

Besides being interesting by themselves, the relations \main\ and
\mainminus\ provide efficient tool for actual computation of the
nonequilibrium expectation values $\langle\, {\bf  V}_{\pm}\, 
\rangle_{BSG}$. 
Using \main,\ \mainminus\ one can easily show that
\eqn\partition{
\langle\, {\bf  V}_{\pm}\, 
\rangle_{BSG}=T\ \partial_\kappa\log
Z_{\pm}(\kappa, V)\ ,
}
where $Z_{+}$ and $Z_{-}$ are the
partition functions in \hibbs\ and \hibbsminus, respectively. On the
other hand, the above partition functions are related in a simple way to
the vacuum eigenvalues $Q_{\pm}(\lambda, p)$ of the CFT analogs of
Baxter's operators ${\bf
Q}_{\pm}(\lambda)$ studied
in\ \refs{\BLZZ, \BLZZZ} (see Sect.4 for some details),
\eqn\zq{
Z_{\pm}(\kappa, V)/Z_\pm(0,V)=
\lambda^{\mp 2\pi i p/\beta^2} Q_{\pm}(\lambda, p)\ .
}
where $Z_\pm(0,V)$ are the partition functions for the ``free''
systems defined by \hplus\ and \hminus\ with $\kappa=0$.
The parameters $\lambda,\ p$\  and $\beta$ used in  
\refs{\BLZZ, \BLZZZ}\ are defined as 
\eqn\mup{
\lambda= i\kappa\ {\sin(\pi g)\over g}\ 
\Big({g\over 2\pi T}\Big)^{1-g}\, ,\qquad
p=-iV\ {g\over 4\pi T}\, ,\qquad \beta^2=g\ .}
Many exact results for these operators exist\ \refs{\BLZZ, \BLZZZ}.
These include high and
low temperature expansions, but most importantly, the
vacuum eigenvalues $Q_{\pm}(\lambda, p)$ are shown in\ \BLZZ\ to satisfy
closed integral equations 
-- the Destri-de Vega equations -- which allow for their evaluation to an
arbitrary degree of accuracy. In view of \main, \mainminus\ and
\partition\ all these results directly apply to the non-equilibrium
expectation values of the operators \vpm.

In applications, the quantity of a particular interest is
\eqn\current{
J =V+ \langle\,  \Phi_{x} (0,t)\, \rangle_{BSG}\ ,
}
which is interpreted as the
current through the point contact in the quantum Hall 
systems\foot{The voltage $V$
and current $J$\ 
\current\ differs 
in normalization from the real voltage
$V^{(phys)}$ and 
current  $J^{(phys)}$\ 
in the Hall system
$$V^{(phys)} = e^{-1}\ V\ ,\ \ \ \ 
J^{(phys)}={e\over2\pi \hbar}\ g\, J\ ,$$
where $e$ and $\hbar$ are the
electron charge  and Plank's constant.
Also, \ $g$ coincides with the
fractional filling of the Luttinger state
in a   Hall bar 
and the temperature $T$ is  measured in  energy units.},
or as the drift velocity in dissipative quantum
mechanics.
It was the main subject of
interest in many recent papers\ \refs{\Kane, \Moon, \FSLSN,\weiss,\SalF,
\BLZZ}.
In particular, for
$g={1\over 3}$ in \bsga, Bethe-Ansatz computation of 
this current was given in\ \FSLSN, under the assumption that the 
Boltzmann equation for the backscattering electrons in the Hall 
contact holds exactly. Further, it was
conjectured in\ \SalF\ that for all values of this parameter
\eqn\their{
J =V+i\pi T\, \mu\,
\partial_{\mu}
\log
\bigg({{Z_{2p}(\mu)}\over{Z_{-2p}(\mu)}}\bigg)\ ,}
with
\eqn\fdre{\mu=\kappa\ {\pi\over g}\
\Big({g\over 2\pi T}\Big)^{1-g}\ , \ \ \  
p=-iV\ {g\over 4\pi T}\ .}
The ``partition function'' $Z_{2p}$ was  defined in Ref.\SalF\ as 
a power series
\eqn\zimp{
Z_{2p}(\mu) =1+\sum_{n=1}^\infty  \mu^{2n}\  I_{2n}(p) 
}
with the coefficients $I_{2n}$ given by multiple infinite sums over
all ordered sets ${\bf m}=(m_1,m_2,\ldots, m_n)$ 
of non-negative integers $m_1\ge
m_2\ge\cdots\ge m_n\ge0$ 
\eqn\itn{
I_{2n}(p)={1\over \Gamma^{2n}(g)}\
\sum_{\bf m}\ 
\prod_{i=1}^n{\Gamma\big(m_i+g(n-i+1)\big)\,
\Gamma\big(2p+m_i+g(n-i+1)\big)\over       
\Gamma\big(m_i+g(n-i)+1\big)\,
\Gamma\big(2p+m_i+g(n-i)+1\big)}\ .}

Similarly looking conjecture
\eqn\our{
J=V+i\pi T\, \kappa\, 
\partial_{\kappa}
\log
\bigg({{Z_{+}(\kappa,V)}\over{Z_{-}(\kappa, V)}}\bigg)\ ,
}
where $Z_{\pm}(\kappa,V)$ are just the partition functions in \hibbs,
\hibbsminus\ and \zq, was proposed independently in\ \BLZZ. The 
(plausible) equivalence between \their\
and \our\ is not yet established. 
However, it is easy to see that \our\ is a
simple consequence of \partition\ and the equation
\eqn\shdyt{\Phi_{x} (0,t) =i \pi \kappa\
\big(\, {\bf{V}}_{+}(t)-{\bf V}_{-}(t)\, \big)\ ,} 
which is the boundary condition corresponding to \bsga, i.e. our result
\main, \mainminus\ actually proves the conjecture \our. 
 
\newsec{Expectation values in boundary sine-Gordon model}

If $V > 0$ the system \bsga\ evolves towards a stationary state
which is characterized by nonzero expectation values $\langle\, \Phi_t
\, \rangle_{BSG}\, , \  \langle\,  \Phi_x\,  \rangle_{BSG}$. This is not a
thermodynamic equilibrium state and no simple explicit
expression for its density matrix is known. In this
section we describe the definition of this state in terms of real-time
perturbation theory and discuss some properties of corresponding
expectation values.

Let us split the total Hamiltonian \bsgh\ into the free and interaction 
parts
\eqn\Hnot{
{\bf H}_0=
{1\over {4\pi g}}\int_{-\infty}^{0}dx\ \big(\, 
\Pi^2 + 
\Phi_x^2\,  \big)\, , 
\qquad {\bf H}_1=-{\kappa\over g}\ \cos(\Phi_B+Vt)\ .}
Using corresponding interaction representation one can write 
the density matrix ${\bf P}(t)$ as
\eqn\dens{
{\bf P}(t)=e^{-i {\bf H}_0  t}\ 
{\bf S}(t,-\infty)\
{\bf P}_0 \ {\bf  S}(-\infty,t)\
e^{i {\bf H}_0  t}\ ,}
where
\eqn\smat{\eqalign{
{\bf S}(t,t_0)={\cal T}& \exp\Big\{-i\int_{t_0}^{t} d\tau\
{\bf H}_1^{(int)}(\tau)\Big\}=\cr
=1+&\sum_{k=1}^\infty\,(-i\/)^k\,
\int_{t_0}^{t} {\cal D}_k(\{\tau\})\  
{\bf H}_1^{(int)}(\tau_1)\, {\bf H}_1^{(int)}(\tau_2) 
\cdots {\bf H}_1^{(int)}(\tau_n)}}
and
\eqn\Hint{
{\bf H}_1^{(int)}(t)=e^{i{\bf H}_0  t}\ {\bf H}_1\
e^{-i{\bf H}_0  t}\ .}
In \smat\ and below the shorthand notation for the multiple ordered
integrals
\eqn\Ddef{
\int_{t_0}^{t}{\cal D}_k(\{\tau\})
=\int_{t_0}^{t}d\tau_1\int_{t_0}^{\tau_1}d\tau_2\cdots
\int_{t_0}^{\tau_{k-1}}d\tau_k}
is used. In writing \dens\ we have assumed that the interaction has 
been adiabatically 
switched on in the infinite  past ($t=-\infty$) when the system was in 
thermodynamic equilibrium at the temperature $T$, i.e.
\eqn\Pnot{
{\bf P}_0=Z_0^{-1}\ e^{-R{\bf H}_0}\, ,\qquad R=g/T \ .}
Then, the expectation value  of an
arbitrary operator ${\bf A}$ can be written
as
\eqn\aver{
\Tr_{\cal F}\big[\, {{\bf  P}(t)\, {\bf A} }\, \big]=
\Tr_{\cal F}\big[\, {{\bf P}_0\,  
{\bf S}(-\infty,t) \,{\bf A}^{(int)}(t)\, 
{\bf S}(t,-\infty)}\, \big]=
\vev{{\, {\bf S}(-\infty,t) \,{\bf A}^{(int)}(t)\,
{\bf S}(t,-\infty)\, }}_0\, ,}
where $\vev{\, \ldots\, }_0$ denotes the expectation value over 
the equilibrium state 
\Pnot\ of the free system, and the superscript 
``$(int)$'' means that this operator is taken in the interaction
representation, i.e. ${\bf A}^{(int)}(t)=
e^{i {\bf H}_0  t}\, {\bf A}\, e^{-i{\bf  H}_0  t}$.
Equivalently, one may write the above expectation value as
\eqn\heiex{
\langle\,  {\bf A}\, \rangle_{BSG}= \Tr_{\cal F} \big[\,
{\bf P}\,  {\bf A}(t)\, \big]\ ,} 
where ${\bf P}$ stands for
density matrix of the system at $t=0$, i.e.
${\bf P}={\bf P}(0)$, and ${\bf A}(t)$ is the
full Heisenberg operator
\eqn\Ahei{
{\bf A}(t)=
{\bf S}(0,t) \,{\bf A}^{(int)}(t)\,
{\bf S}(t,0)\ .}

All above formulae are very well known (see e.g. \Kel); we included them 
here to fix the
notations. In this paper we are interested in the expectation values 
$\vev{\, {\bf V}_\pm\, }_{BSG}$ of the Heisenberg operators
\eqn\vhei{
{\bf V}_\pm(t)={\bf S}(0,t)\, {\bf V}^{(int)}_\pm(t)\, {\bf S}(t,0)\ .}
where 
\eqn\Vpm{
{\bf V}^{(int)}_\pm(t)=
\exp\big\{\pm i \Phi_B^{(int)}(t)\pm iVt\big\}\ ,}
and $\Phi^{(int)}_B(t)$ is the  boundary field $\Phi_B$ in the interaction
representation. It is convenient to introduce also auxiliary
operators
\eqn\vtt{
{\bf V}_\pm (t,t_0)={\bf S}(t_0,t)\,{\bf V}^{(int)}_\pm (t)\,
{\bf S}(t,t_0)\ ,}
where $t_0$ is a parameter. For $t_0 =0$ \vtt\ coincide with the
Heisenberg operators \vhei, and according to \aver\ the expectation
values of \vhei\ can be expressed through \vtt\ as follows
\eqn\vexp{
\vev{\, {\bf V}_\sigma(t)\, }_{BSG}=\lim_{t_{0}\to-\infty}\,\vev{\, 
{\bf V}_\sigma(t,t_{0})\, }_0 \ ,}
where $\sigma=\pm1\, $.
Our nearest goal is to prove the following useful
representation for the operators \vtt\ 
\eqn\vser{\eqalign{
{\bf V}&_\sigma (t, t_0 )=
{\bf V}^{(int)}_\sigma  (t)\times\cr &\Big\{\, 1+
\sum_{k=1}^\infty 
\ \sum_{\s_1,\ldots,\s_k=\pm1} \  C_k(\s|\, \s_1,\ldots,\s_k)\, 
\int_{t_0}^t{\cal D}_k(\{t\})  \ 
{\bf V}^{(int)}_{\s_1}(t_1)\cdots {\bf V}^{(int)}_{\s_k}(t_k)\, \Big\}\ ,}}
where the sum is taken over all arrangements of the 
``charges'' $\s_1,\ldots,\s_k=\pm1$, and the coefficients 
$ C_k(\s|\, \s_1,\ldots,\s_k)$ have the following explicit form
\eqn\Ck{
C_k(\s|\, \s_1,\ldots,\s_k)=
\Big(-{\kappa\over g}\Big)^k\ 
\s_1\cdots\s_k \ q^{-\sum_{j=1}^k\s_j\eta_j}\ 
\prod_{j=1}^k\sin( \pi g\eta_j)\ ,}
with
\eqn\cumcharges{
\eta_j=\sigma+\sum_{s=1
}^{j-1} \s_s\ .}
The fact that this operator can be written as the sum \vser\ with some
coefficients $C_k (\s|\, \s_1, \ldots , \s_k)$ is obvious from 
its definition 
\vtt. Indeed, the series expansions \smat\ for the evolution
operators ${\bf S}$ in \vtt\ allow 
one to represent \vtt\ as a series of multiple
integrals of certain products of ${\bf V}_{\pm}^{(int)}$. 
Then, using the commutation
relations 
\eqn\vv{
{\bf V}_{\s_1}^{(int)}(t_1)\, {\bf V}_{\s_2}^{(int)}(t_2)=
q^{2 \s_1\s_2}\ {\bf V}_{\s_2}^{(int)}(t_2)\, 
{\bf V}_{\s_1}^{(int)}(t_1)\, ,\qquad
t_1>t_2\, ,\qquad q=e^{i\pi g}\ ,}
where $\s_1,\s_2=\pm1$, one can always rewrite each of these integrals
as a combination of the time-ordered integrals as in \vser. 
The easiest way to obtain the coefficients $C_k$ in \vser\ is to notice 
that in view of \smat\ the operators \vtt\ satisfy the 
differential equation
\eqn\vdif{
i\, {\partial\over\partial {t_0}}\, {\bf V}_\s(t,t_0)=
\big[\, {\bf H}^{(int)}_1(t_0)\, ,\,{\bf V}_\s(t,t_0)\, \big]}
with the initial condition
\eqn\vinit{{\bf V}_\s (t,t_0)\big|_{t_0=t}={\bf V}^{(int)}_\s(t)\ .}
It is easy to check that the expansion \vser\ satisfies \vdif\ provided 
the coefficients $C_k$ solve the recurrence relations
\eqn\recC{
C_k(\s|\, \s_1,\ldots,\s_k)={i\, \kappa\over2 g}\ 
(1-q^{-2\s_k\eta_k})\  C_{k-1}(\s|\, \s_1,\ldots,\s_{k-1})\ ,}
where the notation \cumcharges\ is used. With the initial condition
$C_0=1$, which follows from \vinit,\ these relations lead to \Ck. 

A simple consequence of \vser\ is the infinite series representation for
the expectation value \vexp,
\eqn\wf{\langle\, {\bf V}_{\s}\, \rangle_{BSG}=
-{2\pi\s\,  T\over \kappa\, \sin(\pi g)}
\ \sum_{n=1}^\infty \lambda^{2n}
\ \sum_{\s_1,\ldots,\s_{2n-1}} \ \Big(\prod_{j=1}^{2n-1} {\sin
(\pi g \eta_j)\over \sin(\pi g)}\Big)
\ J(\s,\s_1,\ldots,\s_{2n-1}|\, p)\ ,}
where the sum is taken over all arrangements of 
$\s_1,\ldots,\s_{2n-1}=\pm1$
with zero total charge $\s+\sum_{s=1}^{2n-1}\s_s=0$, and 
\eqn\jn{\eqalign{
J_n(\s_0,\s_1&,\ldots,\s_{2n-1}|\, p)=\cr
&\int_{-\infty}^0{\cal D}_{2n-1}(\{\tau\})\   
e^{-2p\sum_{j=1}^{2n-1}\s_j\tau_j}\ \prod_{0\le j< l\le 2n-1} 
\Big(2\sinh\big({\tau_j-\tau_l\over2}\big)\Big)^{2 g\s_j\s_l}}}
with $\tau_0\equiv0$. Here the parameters $\lambda$ and $p$ are
defined by\ \mup.
This representation (more precisely, the corresponding representation
for the current \current) was previously obtained in Ref.\weiss\
by a combinatorial method. The above consideration provides an alternative
derivation of this result. It also promotes the formula \wf\ to
the level of the operator relation \vser\ which can be used in
evaluating the multitime correlation functions.

Let us emphasis here an important feature of the expression \vser\ 
and \wf. The sums over $\s_1,\s_2,\ldots\s_{2n-1}$ there 
exclude configurations where any of the ``cumulative charges''
$\eta_j$,\ $j=1,\ldots,2n-1$\ defined by\ \cumcharges\ vanish, because 
one of the factors $\sin(\pi g\eta_j)$ in \wf\ (as well as in \Ck)
then turn to zero. 
As a result all integrals \jn\ appearing in \wf\ converge at the
lower limit, and that is why the limit $t_0 \to -\infty$ in \vexp\ poses
no difficulty. And of course RHS of\ \vexp,\ \wf\ does
not actually depend on
$t$ as a consequence of the invariance of the 
expectation values upon an overall time shift.

\newsec{The system with $q$-oscillator}

Now we turn to the systems \hplus\ and \hminus\ which involve the
boundary $q$-oscillators, with the aim to derive our main relations \main\
and \mainminus.  These two systems are quite similar and in what
follows only \hplus\ is studied explicitly, and only \main\ is actually
derived; the relation \mainminus\ can be obtained by obvious
modifications of the arguments presented below.

Like in \Hnot, let us split \hplus\ as ${\bf H}_{+} = 
{\bf {\bar H}}_0 + {\bf {\bar
H}}_1$, 
\eqn\Hpnot{
{\bf {\bar H}}_0={\bf H}_0-{V}\,{\bf  h}\, ,
\qquad {\bf {\bar H}}_1=-{\kappa\over 2 g}\
\big(\, {\bf a}_{-}\, e^{i\Phi_B}+{\bf a}_{+}\, e^{-i\Phi_{B}}\, \big)\ ,}
where ${\bf H}_0$ is given by \Hnot\ and the operators ${\bf h}$, 
${\bf a}_\pm$
are defined in \qosc. In the Matsubara representation the equilibrium 
density matrix \hibbs\ reads 
\eqn\matrep{
{\bf  P}_+=Z_+^{-1}(\kappa, V)\ e^{-R\, {\bf {\bar H}}_0}\
{\bf {\bar S}}(-iR,0)\ ,}
where the Matsubara operator ${\bar {\bf S}}(-iR,0)$ is the
``imaginary time'' version of 
the time-evolution operator in the corresponding interaction 
representation,
\eqn\smatplus{
{\bf {\bar S}}(t_2,t_1)={\cal T}\exp\Big\{-i\int_{t_1}^{t_2}dt\ {\bf{\bar
H}}^{(int)}_1(t) \Big\}\, ,\qquad {\bf {\bar
H}}^{(int)}_1(t) =e^{i{\bf {\bar H}}_0 t}\, 
{\bf  {\bar H}}_1\,  e^{-i{\bf {\bar H}}_0 t}\ .}
Note that in view of the commutation relations \qosc\
\eqn\Hint{
{\bf {\bar H}}_{1}^{(int)}(t) =
-{\kappa\over 2\beta^2}\
\big(\, {\bf a}_{-}\, {{\bf V}}_{+}^{(int)}(t)+
{\bf a}_{+}\, {\bf V}_{-}^{(int)}(t)\, \big)\ ,}
where ${\bf V}_{\pm}^{(int)}(t)$ are exactly the operators \Vpm.

We are interested in the expectation values of the operators \wdef\ ,
which can be written as
\eqn\wpmex{
\langle\,  {\bf W}_{\pm}\,  \rangle_{+} = Z_{+}^{-1}\ {\rm
Tr}_{{\cal H}_+}\big[\, e^{-R\, {\bf {\bar H}}_0}\ {\bf {\bar S}}(-iR,0)\
{\bf W}_{\pm}  \, \big]\ .}
The fact that $\langle\, {\bf W}_{+}\, 
\rangle_{+} = \langle\, {\bf  W}_{-}\,  \rangle_{+}$ 
(as stated in \main) is
a particular manifestation of the detailed balance principle for the
equilibrium system \hplus; on formal level it is easily established if
one notices that the commutation relations are invariant with respect to
the transformation ${\bf a}_{+}\to \Lambda\, {\bf a}_{+}, \quad {\bf
a}_{-}\to 
\Lambda^{-1}\,{\bf a}_{-}$,\  where $\Lambda$ is a constant. Therefore
\eqn\wexp{
\vev{{\bf W}_{\pm}}_+=
T\ \partial_{\kappa}\,
\log Z_+ (\kappa, V)\ .} 

It is convenient for our purposes to introduce again an auxiliary time
$t_0$ and rewrite \wpmex\ as 
\eqn\wpmexx{
\langle\,  {\bf W}_{+}\, 
\rangle_{+} = Z_{+}^{-1}\ {\rm
Tr}_{{\cal H}_+}\big[\,
e^{-R{\bf {\bar H}}_0} \ {\bf {\bar S}}(t_0-iR,t_0)\
{\bf W}_{+}(0,t_0)\,   \big]\ ,}
where
\eqn\wtt{
{\bf  W}_{+}(t,t_0)={\bf {\bar S}}(t_0,t)\ 
e^{i{\bf {\bar H}}_0t}\  {\bf W}_{+ }\,  e^{-i{\bf {\bar
H}}_0t}\ 
{\bf {\bar S}}(t,t_0)\ .  }
Our proof of \main\ will be based on remarkably simple representation
for the operator ${\bf W}_{+}(t, t_0)$  similar to \vser. Using
perturbative expansions for the operators ${\bf {\bar S}}$ in \wtt\ with 
the
explicit form \Hint\ of the interaction Hamiltonian, and then applying the
commutation relations \vv\ to achieve the full time ordering for the
operators ${\bf V}_{\pm}^{(int)}$,
one can bring \wtt\ to the form analogous to
\vser, i.e.
\eqn\wser{\eqalign{
&{\bf W}_{ +}(t,t_0)=
{\bf a}_-\, {\bf V}_+^{(int)}(t)\, +\cr
&\ \ {\bf V}_+^{(int)}(t)\ \sum_{k=1}^\infty 
\ \sum_{\s_1,\ldots,\s_k=\pm1} \ {\bf {\bar C}}_k(\s_1,\ldots,\s_k)\
\int_{t_0}^t{\cal D}_k(\{t\})  \ 
{\bf V}_{\s_1}^{(int)}(t_1)\cdots {\bf V}_{\s_k}^{(int)}(t_k)\ ,}}
where this time the coefficients 
${\bf {\bar C}}_k (\s_1, \ldots, \s_k)$ are
not $c$-numbers, 
but some operators acting in $\rho_{+}$. In general, from this
analysis one would expect these coefficients to be complicated
polynomials in the operators ${\bf a}_{+}$ and ${\bf a}_{-}$. 
It turns out, however,
that these coefficients contain
only ${\bf a}_{-}$ but not ${\bf a}_{+}$, and
moreover
\eqn\ckbar{
{\bf {\bar C}}_k(\s_1,\ldots,\s_k)=({\bf a}_-)^{\eta_{k+1}}\ 
C_k(+|\, \s_1,\ldots,\s_k)\ ,}
where $\eta_k$ is given by \cumcharges\ with $\s=+1$, and 
$C_k(\s|\, \s_1,\ldots,\s_k)$ are exactly the  same numerical
coefficients \Ck\ as  
in \vser. The rest of the notations in \wser\ is the same as in \vser. 
The derivation of \ckbar\ is very similar to our proof of \Ck\ in the
previous section. By its definition, the operator \wtt\ must satisfy the
differential equation
\eqn\wdif{
i\, {\partial\over{\partial t_0}}\,{\bf W}_{+}(t,t_0) =
\big[\,{\bf{\bar H}}_1^{(int)}(t_0)\, ,
\, {\bf
W}_{+}(t,t_0)\, \big]\ ,}
with the initial condition
\eqn\winit{
{\bf W}_{+}(t,t_0)\big|_{t_0 = t} = {\bf a}_{-}\, {\bf V}^{(int)}_{+}(t)\ .}
Substituting \wser\ with yet unknown operators ${\bf {\bar  C}}_k$, 
one obtains
the recurrence relations
\eqn\recbar{
{\bf {\bar C}}_k(\s_1,\ldots,\s_k)=
{i\, \kappa\over2 g}\
\big(\, 
{\bf {\bar C}}_{k-1}(\s_1,\ldots,\s_{k-1})\ 
{\bf a}_{-\s_k}-q^{-2\s_k\eta_k}\  
{\bf a}_{-\s_k}\ {\bf {\bar C}}_{k-1}(\s_1,\ldots,\s_{k-1})\, \big)}
with the initial condition $\overline{{\bf C}}_0={\bf a}_-$.
It is straightforward to check that \ckbar\ satisfy \recbar. 
Note that because of \ckbar\ and \Ck, the representation \wser\ enjoys
the same remarkable property as \vser, namely the series \wser\
contains only the terms where no of the ``cumulative charges'' $\eta_j,
\quad j= 1, \ldots, 2k-1$ vanish. We want to stress also that this
simple form of the coefficients ${\bf {\bar C}}_k$ 
is a very special property of
the operator ${\bf W}_{+}(t,t_0)$. Had we taken, say,  ${\bf W}_{-}$ 
instead of ${\bf W}_{+}$ in \wtt, no such simplification in \wser\ 
would occur, i.e. the resulting coefficients ${\bf {\bar C}}_k$
would indeed come out to be complicated polynomials
of both ${\bf a}_{-}$ and
${\bf a}_{+}$, and in particular the above ``no zero cumulative charges''
property would not hold. 

Our proof of \main\ is based on the representation \wser. Consider the
expression \wpmexx\ for $\vev{\, {\bf W}_{+}\, }$ and 
take the limit $t_0 \to
-\infty$. Due to the above ``no zero cumulative charges'' property of
\wser\  the trace in \wpmexx\ in this limit factorizes as
\eqn\factor{\eqalign{
{\rm Tr}_{{\cal H}_+}
\big[\, e^{-R{\bf {\bar H}}_0}\ &
{\bf{\bar  S}}(  t_0-i R, t_0)\ 
{\bf  W}_{+}(t,t_0)\, \big] \to\cr
&{{\rm Tr}_{{\cal H}_+}
\big[\, e^{-R{\bf {\bar H}}_0}\ {\bf W}_{+}(t,t_0)\, \big]
\over {\rm Tr}_{{\cal H}_+}
\big[\,e^{-R{\bf {\bar H}}_0}\, \big]}\ \ 
{\rm Tr}_{{\cal H}_+}
\big[\, e^{-R{{\bf \bar H}}_0}\ {\bf{\bar  S}}(t_0-iR ,  t_0)\, \big]\ ,}}
and we obtain
\eqn\wlim{
\vev{\, {\bf W}_{+}\, }_{+}=
\lim_{t_0 \to -\infty}\ 
{{\rm Tr}_{{\cal H}_+}
\big[\, e^{-R{\bf {\bar H}}_0}\ {\bf W}_{+}(0,t_0)\, \big]\over 
{\rm Tr}_{{\cal H}_+}
\big[\,e^{-R{\bf {\bar H}}_0}\, \big]}\ .}
The trace in the denominator is the partition
function of\ \hplus\ with $\kappa=0.$
Now, using here the representation \wser\
with \ckbar\ for the operator ${\bf  W}_{+}(0,t_0)$, and taking
advantage of the fact that 
\eqn\aav{
{\rm Tr}_{{\cal H}_+}
\big[\,  e^{-R{\bf {\bar H}}_0}\, 
({\bf a}_{-})^n\,\, \big] =
\delta_{n,0} \ {\rm Tr}_{{\cal H}_+}
\big[\,e^{-R{\bf {\bar H}}_0}\, \big]\ ,}
one arrives exactly at the series in \wf\ for
the expectation value \wlim. This proves the relation \main. Let us
also note that using the same arguments based on the properties of 
\vser,\ \wser\ and on \aav, one can establish 
more general relation between the multitime correlation functions of the
systems \bsgh\ and \hplus, namely
\eqn\multi{
\vev{\, {\bf V}_{+}(t_1)\,\cdots\,{\bf  V}_{+}(t_n)\, }_{BSG} =
\vev{\, {\bf W}_{+}(t_1)\,\cdots\,{\bf W}_{+}(t_n)\, }_{+}\ .}
Here ${\bf  W}_{+}(t)={\bf  W}_{+}(t,0)$ are 
the Heisenberg operators
associated with ${\bf W}_{+}$.

\newsec{Discussion}

The relations \main\ and \mainminus\ between nonequilibrium expectation
values in \bsga\ and equilibrium expectation values in \hplus\ and
\hminus\ is the main result of this paper. We want to stress here that
these relations are actually derived (in Sect.2 and
3) from first principles, and in particular this result proves the validity
of the expression \our\ for the current \current\
conjectured earlier in \BLZZ.
General significance of this result depends on
whether it can be extended to wider class of systems with nonequilibrium
dynamics. At this time we feel reluctant to enter any speculations on
this matter leaving that to when better understanding of the situation
is achieved. Instead, in this section we discuss how these relations can
be applied towards actual computation of the nonequilibrium expectations
$\langle\, {\bf V}_{\pm}\, \rangle_{BSG}$ and of the current 
\current.

According to \partition\ and \our, computation of these expectation
values reduces to finding the partition functions $Z_{\pm}(\kappa, V)$
of the equilibrium systems \hplus\ and \hminus. By itself this does
not help much in solving the problem. Of course one can use Matsubara
theory to write down a perturbative series in $\kappa^2$ for
$Z_{\pm}(\kappa,V)$. However, $n$-th order coefficients of this series are
expressed in terms of $2n-1$ fold integrals (written down explicitly in
\BLZZ) and it is not clear how to evaluate or simplify these integrals
in general. In this respect this approach does not have any significant
advantages over the series \wf; in
fact, \wf\ looks more compact. The same remark seems to apply to the
approach based on the representation \their\ conjectured in \SalF.
Even if \their\ is proven, evaluation of all terms of the series \zimp\
requires calculation of the sums \itn\ of ever growing degree of
complexity. Each of these representations --- \our, \wf\ and \their\ ---
provides more or less efficient way to calculate few first terms of 
the $\kappa^2$ expansion of \current\ but neither seem to give a full
solution to the problem. In fact, if one is concerned with the actual
evaluation of the coefficients of these $\kappa^2$ expansions, even more
efficient technique can be obtained by combining certain analyticity
properties and functional relations for ${\bf Q}$ operators (see
below); we describe this technique in the Appendix.

What makes the relations \main, \mainminus\ a useful computational
tool is remarkable relation \zq\ between the partition functions
$Z_{\pm}(\kappa, V)$ and certain eigenvalues of so-called ${\bf
Q}_{\pm}$-operators. These operators were introduced in \BLZZ\ as CFT 
analogue of Baxter's Q-matrix \Baxn. The operators ${\bf
Q}_{\pm}(\lambda)$ of \BLZZ\ act in the Fock space of free Bose field 
with spatial coordinate compactified on a circle. The Fock vacua 
$\mid p\, \rangle$ are parameterized by the value $p$ of the zero-mode 
momentum. These vacua are eigenstates of the operators ${\bf Q}_{\pm}
(\lambda)$, and we denote $Q_{\pm}(\lambda, p)$ the corresponding
eigenvalues, i.e.
\eqn\eigenv{
Q_{\pm}(\lambda, p) = \langle\, p \mid {\bf Q}_{\pm}(\lambda) \mid p\,
\rangle \ .}
We will also use the notations
\eqn\apm{
A_{\pm}(\lambda, p) = \lambda^{\mp{2\pi i p}/
\beta^2}\,Q_{\pm}(\lambda, p) \ .}
The identity \zq, i.e.
\eqn\adef{Z_{\pm}(\kappa, V)/Z_\pm(0,V)=A_{\pm}(\lambda,p)\ , }
(where the parameters $\kappa,\, V,\, g$  and $\lambda,
\, p,\, \beta^2$ are  related by \mup) can be verified by direct
comparison of the definition of $Z_{\pm}(\kappa, V)$ in Sect.1 and the
definition of the ${\bf Q}$-operators in \BLZZ. At the same time the
above operators ${\bf Q}_{\pm}(\lambda)$ exhibit some very useful
properties. First, the operators 
$\lambda^{\mp2\pi i P / \beta^2}\,{\bf Q}_{\pm}(\lambda)$, 
and hence the functions \eigenv, 
are entire functions of $\lambda^2$,
with known asymptotics at $\lambda^2 \to -\infty$ along the real axis
\BLZZ. Second, these operators obey certain functional equations (one
is utilized in Appendix, see (A.5)),
notably the famous Baxter's $T$-$Q$ relation. These analytic
characteristics and functional equations lead to closed nonlinear
integral equation for the functions \adef\ (explicitly written down in 
\BLZZ\ \foot{In the form written down in \BLZZ\ the Destri-de Vega
equation applies only to the case of real $2p > -g$, i.e. pure imaginary
$V$. For generic complex values of $p$ the equation must be modified,
namely the $\lambda$ integral over the real axis must be replaced by
contour integral with the contour encircling all zeroes of
$A_{\pm}(\lambda, p)$ in the $\lambda$ plane, which become complex at
complex $p$.}), known as Destri-de Vega equation \refs{\DDV,\KBP}. 
This equation
has been used in \BLZZ\ to analyze low-temperature expansions for the
current \current\ and to confirm notable duality relation for this
current first suggested in \Fisher\ (some relevant results
concerning this duality were obtained in \refs{\Schmid,\FSLSN,\Wei}). 
Detailed study
of the solutions to this equation lies beyond the scope of this work.

\bigskip

\centerline{ {\bf Acknowledgments}}

\bigskip

V.B.  thanks
members of RIMS,
Kyoto University, for their kind hospitality during the final stages 
of this work.
S.L. acknowledges kind hospitality of SdPT at Saclay
where parts of this work were done. A.Z. is grateful to
Al. Zamolodchikov and G. Falkovich for encouraging interest to this work
and many discussions.  
Research of S.L. and A.Z.
is supported in part by DOE grant \#DE-FG05-90ER40559.

\appendix{A}{}

In this section we describe a technique of calculating the perturbative
expansions in $\kappa^2$ (``high
temperature expansions'')
of the expectation values \main, \mainminus\
and \current. The technique is based on the functional relations for
the ${\bf Q}$-operators derived in \refs{\BLZZ, \BLZZZ}. Unlike the main
text of the paper which was meant to be more or less self-contained, this
Appendix relies heavily on the results of Refs.\refs{\BLZZ, \BLZZZ}, so
some familiarity with these works will be useful in reading it.

Let us remind some properties of the functions \apm\ which follow from
the analysis in \refs{\BLZZ,\BLZZZ}.

(i) The functions $A_{\pm}(\lambda ,p)$ are entire functions of
$\lambda^2$, and $A_{\pm}(0,p)=1$. The power series \foot{
The coefficients $a_n(p)$ here differ in normalization 
from $H_n^{(vac)}(p)$ 
used in \BLZZ, namely
$$a_n(p)=g^{-2 n}\ \Gamma^{2n}(1-g)\ H_n^{(vac)}(p)\ .$$ } ,
\eqn\logA{\log A_{+}(\lambda, p)=
-\sum_{n=1}^{\infty}\,  a_n(p)\ \lambda^{2 n}}
has finite radius of convergence which is determined by position of
closest to origin zero of $A_{+}(\lambda, p)$. For real $2p>-g$ this
zero is a  real  and positive. Let us mention here that according to
\partition
, \wf\ and \adef, the coefficient $a_n(p)$ can be expressed in terms
of the integrals \ \jn,
\eqn\anint{a_n(p)={\pi\over n\, \sin(\pi g)}\, \
\sum_{\s_1+\ldots+\s_{2n-1}=-1} \ \Big(\prod_{j=1}^{2n-1} {\sin
(\pi g\eta_j)\over \sin(\pi g)}\Big)\  
\ J(+1,\s_1,\ldots,\s_{2n-1}|\, p)\ ,}
with $\sigma_s=\pm 1,\ \eta_j=1+\sum_{s=1
}^{j-1} \s_s$.
Note that this relation holds {\it as is} only when $0<g<{1\over 2}$ and 
$ \Re e\, (2p)>-g$, since otherwise the integrals \jn\ generally
diverge; analytic continuation in these parameters is required outside
this domain. 

(ii) The coefficients  $a_n(p)$ are meromorphic functions of $p$.
Moreover, they are  analytic in the half plane 
$\Re e\, (2p)>-g$. The last statement  follows
directly from Eqs.\jn\ and \anint.

(iii) The following asymptotics,
\eqn\osoeuiu{a_n(p)\to {\Gamma(n g)\, \Gamma\big(-{1\over 2}+
n (1-g)\big) \over 2\, \sqrt{\pi}\, n!}\ \Gamma^{2 n} (1-g)\ 
p^{1-2 n+2 n g }\, , \ \ \ \ \  {\rm as}\  \ p\to\infty \ ,}
hold in the half-plane $\Re e\,(2p)>-g$.

(iv) $A_-(\lambda, p)$ is related to
$A_+(\lambda, p)$  as
\eqn\hsdgtsd{A_{-}(\lambda, p)=A_{+}(\lambda, -p)=
\exp\Big\{-\sum_{n=1}^{\infty}\, a_n(-p)\, \lambda^{2 n}\, \Big\}\,  .}

(v) The functions $A_{\pm}(\lambda, p)$ obey so called
``quantum Wronskian'' condition,
\eqn\wrons{e^{2\pi i p}\  
A_{+}(q^{1\over 2}\lambda, p)\, A_{-}(q^{-{1\over 2}}\lambda, p)-
e^{-2\pi i p}\ A_{+}(q^{-{1\over 2}} \lambda, p)\,
A_{-}(q^{{1\over 2}}\lambda, p)=2 i\, \sin(2\pi p)\ ,}  
where $q=e^{i\pi g}$.

The Eqs.\hsdgtsd\ and \wrons, supplemented with the analyticity (ii)
and asymptotic conditions \osoeuiu\ constitute
a Riemann-Hilbert problem which
defines the functions $A_{\pm} (\lambda, p)$\ 
completely. Indeed, substituting the expansions \logA, \hsdgtsd\ 
into the \wrons,\ one  obtains relations of the form      
\eqn\jsdgtr{\sin(\pi n g+2\pi p)\ 
a_n(p)-\sin(\pi n g-2\pi p)\ 
a_n(-p)=R_n(p)\ ,\ \ \ n=1,2,\ldots\ , }
where the functions  $R_n(p)$
are expressed through $a_k(p)$ with $k=1,\ldots n-1$ only.
For example,
\eqn\skjdyt{\eqalign{&R_1(p)=0\, , \cr
&R_2(p)=-\big( \, q\, a_1(p)+q^{-1}\, a_1(-p)\, \big)^2\ 
e^{4\pi i p}\ \sin(2\pi p)/2\ .}} 
Since $a_n(p)$ are analytic in the half plane
$\Re e\,  (2 p)>-g$, and $a_n(p)\to
const\  p^{1-2 n+2 n g}$\ as $p\to\infty$ there, one can solve 
the Eq.\jsdgtr\ with respect to $a_n(p)$,
\eqn\hsgdtr{\eqalign{a_1(p)=&{\pi
\Gamma(1-2 g)\over \sin(\pi g)}\ 
{\Gamma(g+2 p)
\over \Gamma(1-g+2 p)}\, ,\cr
a_n(p)=&(-1)^n\  {i\over \pi}\ {\Gamma(2-n+n g+2 p)
\over \Gamma(n-1-n g+2 p)}\times\cr
& \int_{-\infty}^{+\infty}\, 
{d x\over 2\pi}\ {R_n(i x)\ \Gamma(n-1-n g+2 i x)\, 
\Gamma(n-1-n g -2 i x)\over x+i p}\ ,}}
where $n=2,3,\ldots\ . $ In writing\ \hsgdtr,\ we assume 
that $ 0<g <{1\over 2}$ and $\Re  e\, p>0$. 
As the functions $R_n(p)$ are uniquely expressed (by the use of \skjdyt
) through ``lower'' $a_k(p)$ with $k=1,2,\ldots ,n-1$, the Eq.\hsgdtr\ 
provides a recursion for evaluation of $a_n(p)$. It allows one to represent
$a_n(p)$ in terms of $(n-1)$--fold  integral.
Let us present  explicit formulae for $a_2$ and $a_3$\ 
($ 0<g <{1\over 2}$ and $\Re  e\, p>0$),
\eqn\heyrtr{\eqalign{
&a_2(p)= 2^{1-4 g}\ {\Gamma^2(1-g )\over \Gamma^2({1\over 2}+
g)}\ 
{\Gamma(2g+2 p)\over \Gamma(1-2g +2 p)}\ 
\int_{-\infty}^{+\infty}{d x\over 2 \pi}   
\ {S_1(x)\over x+i p}\ ,\cr
&a_3(p)= 2^{2-6 g}\ \sqrt{\pi}\ {\Gamma^3(1-g)\over
\Gamma^3({1\over 2}+
g)}\ 
{\Gamma(3g-1+2 p)
\over  \Gamma(2-3g +2 p)}\times\cr& \biggl\{\,
-{\sin(4\pi g)\over \pi^2}\ 
\int_{-\infty}^{+\infty}{d y\over 2 \pi}
\int_{-\infty}^{+\infty}{d x\over 2 \pi}\
{S_2(x,y)\over (y+i p )(x-y-i 0)}\ 
+{1\over 3}\, \int_{-\infty}^{+\infty}{d x\over 2 \pi} 
\ {S_3(x)\over x+i p}\ \biggl\}\ ,}}
where  the functions $S$ are
\eqn\rewwt{\eqalign{&S_1(x)=\sinh(2 \pi x)\, \Gamma(1-2 g+2 ix)
\, \Gamma(1-2 g-2 ix)\,  
\big(\Gamma(g+2 i x)\, \Gamma(g-2 i x)\big)^2\ ,\cr
&S_2(x,y)=\sinh(2 \pi y)\, \sinh(2 \pi x)\, \Gamma(g+2 i y)\,
\Gamma(g-2 i y)\,
\Gamma(2g+2 i y)\, \Gamma(2g-2 i y)\times\cr
&\ \ \ \ \ \ \ \ \ \ \Gamma(2-3g+2 i y)\, \Gamma(2-3g-2 i y)\,
\Gamma(1-2g+2 i x)\, \Gamma(1-2g-2 i x)\times\cr&
\ \ \ \ \ \ \ \ \ \ \
\big(\Gamma(g+2 i x)\, \Gamma(g-2 i x)\big)^2\  ,\cr
&S_3(x)=\sinh(2 \pi x)\, \Gamma(2-3 g+2 ix)
\, \Gamma(2-3 g-2 ix)\, 
\big(\Gamma(g+2 i x)\, \Gamma(g-2 i x)\big)^3\times\cr & 
\ \ \ \ \ \ \ \ \ \ \ {\sin(4i\pi x+2\pi g)-2 \sin(2\pi g)\over  
\sin(2i\pi x+2\pi g)}\ .}}
Eqs.\hsgdtr,\ \heyrtr\ have obvious computational advantages over
\anint. First, the number of integrations is greatly reduced - while
\anint\ requires evaluation of $2n-1$-fold integral \jn\ to compute
$a_n(p)$, the recursion relation \hsgdtr\ leads to $n-1$ fold integral
for this quantity. Second, it is very convenient in numerical
calculations because the integral in \hsgdtr\ converges very fast at
infinity. It is important to note also that analytic
continuation of the expressions\ \hsgdtr,\ \heyrtr\ outside the domain 
$ 0<g <{1\over 2}$ and $\Re  e\, p>0$ is rather straightforward --- it
is done by appropriate deformation  of integration contours.
For example, the analytical continuation of
$a_2(p)$ into the domain ${1\over 2} < g<1$ and $\Re  e\, p>0$ reads,
\eqn\hetr{\eqalign{
a_2(p)= &2^{1-4 g}\ {\Gamma^2(1-g )\over \Gamma^2({1\over 2}+
g)}\
{\Gamma(2g+2 p)\over \Gamma(1-2g+2 p)}\ \bigg\{
\int_{-\infty}^{+\infty}{d x\over 2 \pi}
\ {S_1(x)\over x+i p}-\cr
&{\sin(2 \pi g)\, \Gamma(3-4 g)\, 
\Gamma^2(1-g)\, \Gamma^2(3 g-1)\over
(2 p+1-2 g)\, (2 p-1+2 g)}\ \bigg\}\ , }}
where the function $S_1(x)$ is the same as in\ \rewwt.

\listrefs

\end